\documentclass[%
 reprint,
 amsmath,amssymb,
 aps,
]{revtex4-2}
\usepackage{graphicx}
\usepackage{dcolumn}
\usepackage{bm}
\usepackage[usenames]{color}
\usepackage{colortbl}
\usepackage{hyperref}
\begin{document}

\title{Holographic reconstruction of magnetic field distribution in a Josephson junction from diffraction-like $I_c(H)$ patterns}

\author{Razmik A. Hovhannisyan}
\author{Taras Golod}
\author{Vladimir M. Krasnov}
\email{vladimir.krasnov@fysik.su.se} 

\affiliation{Department
of Physics, Stockholm University, AlbaNova University Center,
SE-10691 Stockholm, Sweden} 



\begin{abstract}
A general problem of magnetic sensors is a trade-off between spatial resolution and magnetic field sensitivity. With decreasing sensor size its resolution is improved but the sensitivity is deteriorated. Obviation of such the trade-off requires development of super-resolution imaging technique, not limited by the sensor size. Here we present a proof of concept for a super-resolution method of magnetic imaging by a Josephson junction. It is based on a solution of an inverse problem - reconstruction of a local magnetic field distribution within a junction from the dependence of the critical current on an external magnetic field,  $I_c(H)$. The method resembles the Fourier-transform holography, with the diffraction-like $I_c(H)$ pattern serving as a hologram. A simple inverse problem solution, valid for an arbitrary symmetric case, is derived. We verify the method numerically and show that the accuracy of reconstruction does not depend on the junction size and is only limited by the field range of the $I_c(H)$ pattern. Finally, the method is tested experimentally using planar Nb Josephson junctions. Super-resolution reconstruction of stray magnetic fields from an Abrikosov vortex, trapped in the junction electrodes, is demonstrated. Thus, our method facilitate both the high field sensitivity and high spatial resolution, obviating the trade-off problem of magnetic sensors. We conclude that the holographic magnetic imaging by a planar Josephson junction can be used in scanning probe microscopy.
\end{abstract}

\maketitle

\section{Introduction}

Magnetic scanning probe microscopy (SPM) has been rapidly developing in recent decades. Magnetic force (MFM) \cite{Hug,Volodin,Bud,Bud2,Bud3,LocalJVgen,grebenchuk2020observation}, Superconducting Quantum Interference Device (SQUID) \cite{Kirtley_1999Rev,Moler_2008,Bouchiat_2009Rev,Granata_2016Rev,Kirtley_2016,Zeldov_2013,Zeldov_2016,Tzalenchuk_2003,Zeldov_2017}, Hall-probe \cite{Grigorenko_2001,Kalisky_2009} and NV-center \cite{Maletinsky_2012,Rondin_2012,Sun_2021,Tetienne_2016,Barbiero_2020} microscopies achieved remarkable advances. However, many magnetic sensors suffer from the trade-off problem between spatial resolution and magnetic field sensitivity. For example, SQUIDs   
detect a fraction of the flux quantum, $\Phi_0$ 
\cite{Koelle_1999Rev,Kirtley_1999Rev}. Therefore, their field sensitivity is inversely proportional to the pickup loop area, while spatial resolution is determined by the loop size. 
Thus, miniaturization leads to the improvement of the resolution at the expense of sensitivity. 

In Ref. \cite{VanHarlingen_1999} it was proposed to use a single sandwich-type Josephson junction (JJ) as an SPM sensor. This enables ultimate miniaturization and improves spatial resolution \cite{Zeldov_2013}, but the trade-off problem persists. In Ref. \cite{Golod_2019a} it was argued that planar JJs \cite{Kogan_2001,Boris_2013} would allow at least partial obviation of the problem. 
Josephson effect appears as a result of electronic wave function interference between two superconducting electrodes \cite{barone1982physics}. It leads to diffraction-like Fraunhofer modulation, $I_c(H)$, of the critical current as a function of magnetic field. A local magnetic field, $H^*(x)$, induced in a JJ by a small magnetic object, leads to a distortion of the $I_c(H)$ pattern. In Ref. \cite{krasnov2020josephson} it was argued that $H^*(x)$ is encoded in the shape of $I_c(H)$ and that a restoration of this information would allow super-resolution imaging not limited by the JJ size. This requires solution of an inverse problem - reconstruction of unknown $H^*(x)$ from the known $I_c(H)$.

In this work we show both theoretically and experimentally that 
magnetic field distribution in a JJ can be reconstructed via inverse Fourier transform of the $I_c(H)$ pattern. The method resembles the Fourier-transform holography \cite{Bruckstein_1997,Latychevskaia_2015,Vetal_2018}, with a diffraction-like $I_c(H)$ pattern serving as a hologram. 
A simple solution, valid for an arbitrary symmetric case, is derived. We verify the method numerically and show that the accuracy of reconstruction does not depend on the junction size and is only limited by the field range of the $I_c(H)$ pattern. Finally, the method is tested experimentally using planar Nb Josephson junctions. Super-resolution reconstruction of stray magnetic fields from an Abrikosov vortex, trapped in the junction electrodes, is demonstrated. We conclude that the holographic imaging by planar JJs facilitates both high field sensitivity and high spatial resolution, thus obviating the trade-off problem in SPM. 
 
The paper is organized as follows. First, we present the inverse problem solution, allowing accurate reconstruction of the local field within the junction via inverse Fourier transform of the $I_c(H)$ pattern. The solution is valid for any, symmetric with respect to the junction center, $x=0$, local field $H^*(x)$. Next, we verify the solution numerically for various local field distributions and analyze the accuracy of reconstruction as a function of the flux range of the $I_c(\Phi)$ pattern. It is shown that the accuracy rapidly improves with increasing the number of lobes in the $I_c(\Phi)$ pattern and that the maximum flux range, $\Phi_{max} \sim 5-10~ \Phi_0$, is sufficient for a quantitatively correct reconstruction. Finally, we verify the method experimentally, using planar Nb junctions. We demonstrate a successful reconstruction of stray magnetic fields from an Abrikosov vortex, trapped in the junction electrodes. The super-resolution ability to detect a local field, $H^*(x)$, with a spatial resolution much better than the junction length is, therefore, confirmed. In the Appendix, we provide additional clarifications about the inverse problem solution and describe image improvements by means of analytic continuation and Fourier filtering.  
 
\section{Results and discussion}

When a JJ is placed near a small magnetic object, the object generates a local inhomogeneous magnetic field, $H^*(x)$, within the junction. This distorts the $I_c(H)$ pattern. The corresponding direct problem, i.e., calculation of $I_c(H)$ for a given $H^*(x)$, has been solved in Ref. \cite{krasnov2020josephson}. Here we focus on a solution of the inverse problem - reconstruction of the unknown local field within the junction from the known $I_c(H)$ pattern.

\subsection{Inverse problem solution}

We consider a ``short" JJ with the length $L< 4\lambda_J$, where $\lambda_J$ is the Josephson penetration depth. In this case we can neglect screening effects and other complications, associated with Josephson vortices.  
The field $H_y$ (in the $y$-axis direction) induces a gradient of Josephson phase shift along the junction (in the $x$-axis direction), 
$d\varphi /d x = \alpha H_y$, 
, where $\alpha=2\pi d_{eff}/\Phi_0$ and $d_{eff}$ is the effective magnetic thickness of the JJ. The homogeneous external field $H$ creates a constant gradient, but 
a small magnetic object with a Gaussian-like local field, $H^*(x)$,
creates a step-like phase shift,
\begin{equation}\label{psi*}
\varphi^* (x) = \alpha \int\limits_{0}^{x} H^*(\xi)d \xi. ~~~~
\end{equation}
$I_c$ is obtained by maximization of the Josephson current,   
\begin{equation}\label{I_c}
I_c  = \int\limits_{-L/2}^{L/2}{J_c(x)\sin \left[ \alpha H x + \varphi^*(x) +\varphi_0\right] dx},~~~~
\end{equation}
with respect to $\varphi_0$. Here $J_c(x)$ is the critical current density, 
which may vary along the JJ \cite{Dynes_1971,Krasnov_1997}. 

To solve the inverse problem we must extract $\varphi^*(x)$ from a given $I_c(H)$. 
First, we note that 

\begin{equation}\label{i I_c}
i I_c(h)  = \int\limits_{-L/2}^{L/2}{J_c(x)\exp \left[ i( h x + \varphi^*(x) +\varphi_0(h) ) \right] dx},
\end{equation}
where $h = \alpha H$. This follows from the Eulers formula taking into account that 

\begin{equation}
    \int\limits_{-L/2}^{L/2} J_c(\xi)\cos(h\xi + \varphi^* (\xi)+ \varphi_0(h))d\xi = 0,
\end{equation}
due to the maximization requirement $\partial I_c/\partial \varphi_0 = 0$. Considering Eq. (\ref{i I_c}) as the direct Fourier integral, we can perform the inverse Fourier transform for solely $x$-dependent term $J_c(x)e^{i\varphi^*(x)}$: 

\begin{equation}
\label{FFU}
   \begin{split}
    J_c(x)e^{i\varphi^*(x)} = \frac{i}{2\pi} \int\limits_{-\infty}^{\infty} e^{-ixh}e^{-i\varphi_0(h)}I_c(h) dh.  ~~~
    \end{split}
\end{equation}
Imaginary and real parts of Eq. (\ref{FFU}) lead to a system of two equations for $\varphi^*(x)$:
\begin{eqnarray}
J_c(x)\sin[\varphi^*(x)] = \frac{\alpha}{2\pi} \int\limits_{-\infty}^\infty \cos [\alpha x H + \varphi_0(H)] I_c(H)dH,~~~~\label{eqn6}\\
J_c(x)\cos[\varphi^*(x)] = \frac{\alpha}{2\pi} \int\limits_{-\infty}^\infty \sin [\alpha x H + \varphi_0(H)] I_c(H) dH.~~~~\label{eqn7}
\end{eqnarray} 
The unknown $\varphi_0(H)$ should be obtained from the extremum condition, $\partial I_c/\partial \varphi_0=0$, which yields:  
\begin{equation}\label{f0}
    \varphi_0(H) = \frac{\pi}{2} - \arctan\left[\frac{A(H)}{B(H)}\right],
\end{equation}
where $A(H) = \int_{-L/2}^{L/2}  J_c(x) \sin \left[ \alpha H x + \varphi^*(x) \right]dx$ and $B(H) = \int_{-L/2}^{L/2}  J_c(x) \cos \left[ \alpha H x + \varphi^*(x) \right]dx$.

\begin{figure*}[t]
    \begin{center}
        \includegraphics[width=0.95\textwidth]{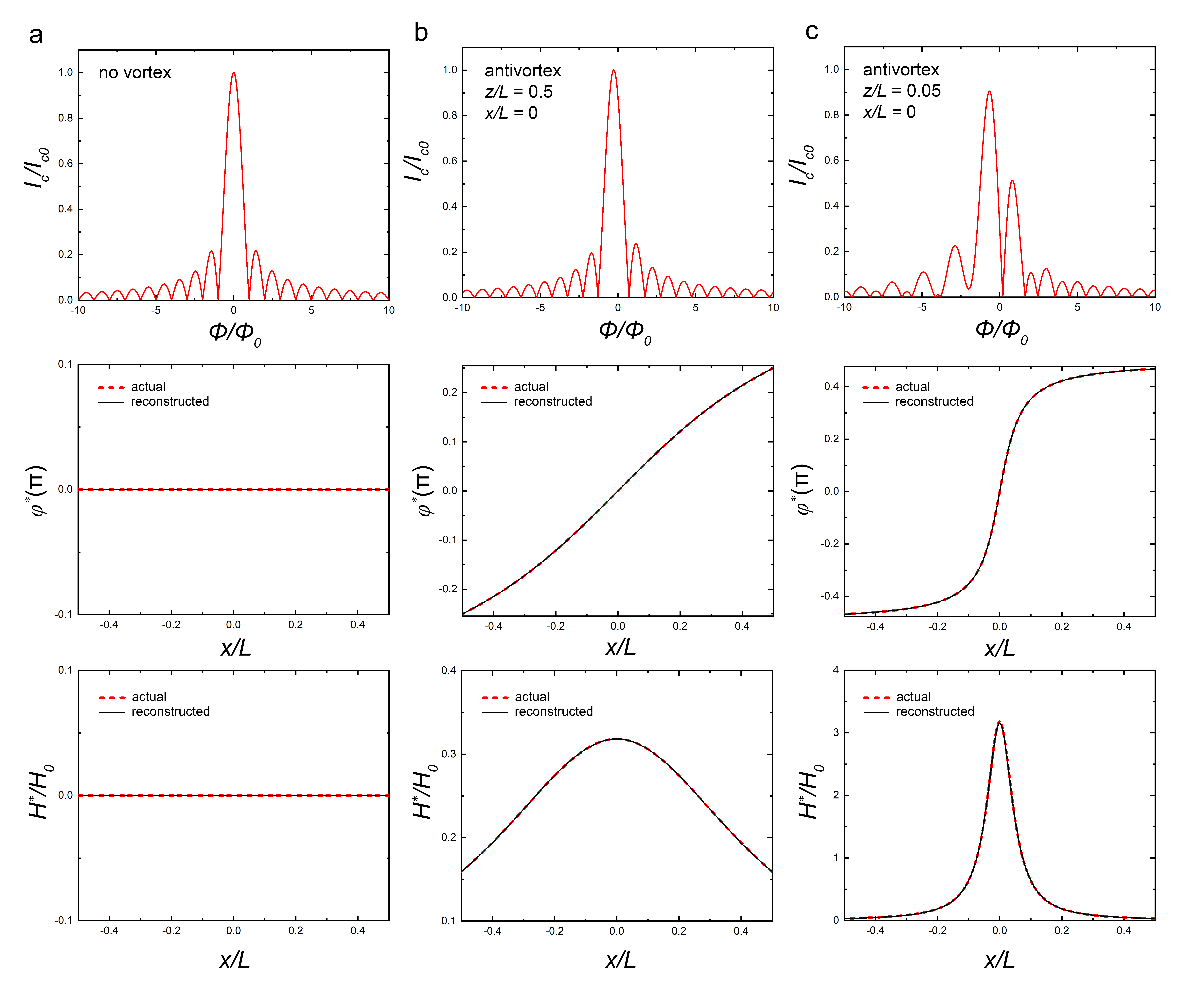}
        \caption{Calculated characteristics (a) without a vortex ($z_v=\infty$), and (b,c) with local stray fields from an antivortex placed (b) at a moderate distance, $z_v = 0.5~ L$ and (c) close to the JJ, $z_v = 0.05~ L$. Top panels represent the 
        $I_c$ versus flux modulation; middle panels - the local phase
shifts $\varphi^*(x)$; and bottom panels - the local magnetic fields, $H^*(x)$, normalized by the flux quantization field $H_0$. Red lines represent actual characteristics obtained from Eq. (\ref{AQ1}). Black lines in the middle and bottom panels represent $\varphi^*(x)$ and $H^*(x)$ obtained by the holographic reconstruction of the $I_c(H)$ patterns from the top panels.}
        \label{fig2}
    \end{center}
\end{figure*}

In the absence of the object, $\varphi^*=0$, for a uniform JJ, $J_c(x)=I_{c0}/L$, the term $A$ vanishes because the integrand is odd in $x$. In this case, Eq.~(\ref{f0}) yields $\varphi_0=\pi/2$ and  $I_c(f)$ exhibits Fraunhofer modulation, $I_{c0}\sin(\pi f)/\pi f $, where $f= \Phi/\Phi_0=H/H_0$ is the normalized flux and $H_0$ is the flux quantization field. Substitution of $\varphi_0=\pi/2$ and the Fraunhofer $I_c(H)$ in Eqs. (\ref{eqn6},~\ref{eqn7})
leads to $\sin(\varphi^*)=0$, $\cos(\varphi^*)=1$, verifying reconstruction of the trivial case, as shown in Fig. \ref{fig2} (a). 

It should be noted that the condition $\partial I_c/\partial \varphi_0=0$, used in derivation of Eq.~(\ref{FFU}), defines an extremum, i.e., it provides either a minimum or a maximum. Direct substitution of $\phi_0=\pi/2$ in Eq.~(\ref{I_c}) for the trivial case $\varphi^*=0$, yields $I_{c0}\sin(\pi f)/\pi f $. It changes sign at every consecutive lobe. I.e., if at one lobe it corresponds to the maximum, $I_{c+}$, at the next it would be the minimum, $I_{c-}$. For short JJs they are always correlated $I_{c+}(H) = -I_{c-}(H)$, as can be seen from Fig.~3 in Ref. \cite{krasnov2020josephson}. Therefore, in order to apply the inverse Fourier transform to the experimental $I_{c+}(H)$ pattern, it should be first prepared by flipping the sign at odd lobes:
\begin{equation}\label{sign}
    I_c(H) = I_{c+}(H)(-1)^{n},
\end{equation}
where $n$ is the lobe number counted in both directions from the central, $n=0$, lobe. More discussion about sign alternation procedure can be found in Appendix A.

For $H^*\ne 0$, $\varphi_0$ may depend both on $H$ and $H^*$, preventing a straightforward solution. As usual, the inverse problem requires additional knowledge about the object. In SPM we are primarily interested in imaging of small magnetic objects, such as vortices or domain walls, with spatially symmetric $H^*(x)$. When a symmetric object is placed in the middle of a JJ with a symmetric $J_c(x)$, the term $A(H)$ in Eq. (\ref{f0}) vanishes again, so that $\varphi_0 = \pi/2$ and the inverse solution, Eqs. (\ref{eqn6},\ref{eqn7}), remains unambiguous. 
The most accurate reconstruction is achieved using $\tan[\varphi^*]$ obtained by solving both Eqs.~\ref{eqn6} and \ref{eqn7}. Mutual division of Eqs.~\ref{eqn6} and \ref{eqn7} eliminates the $J_c(x)$ term. This is important for practical application when $J_c(x)$ is not confidently known. All solutions presented below are obtained this way. 

\begin{figure*}[t]
        \includegraphics[width=0.95\textwidth]{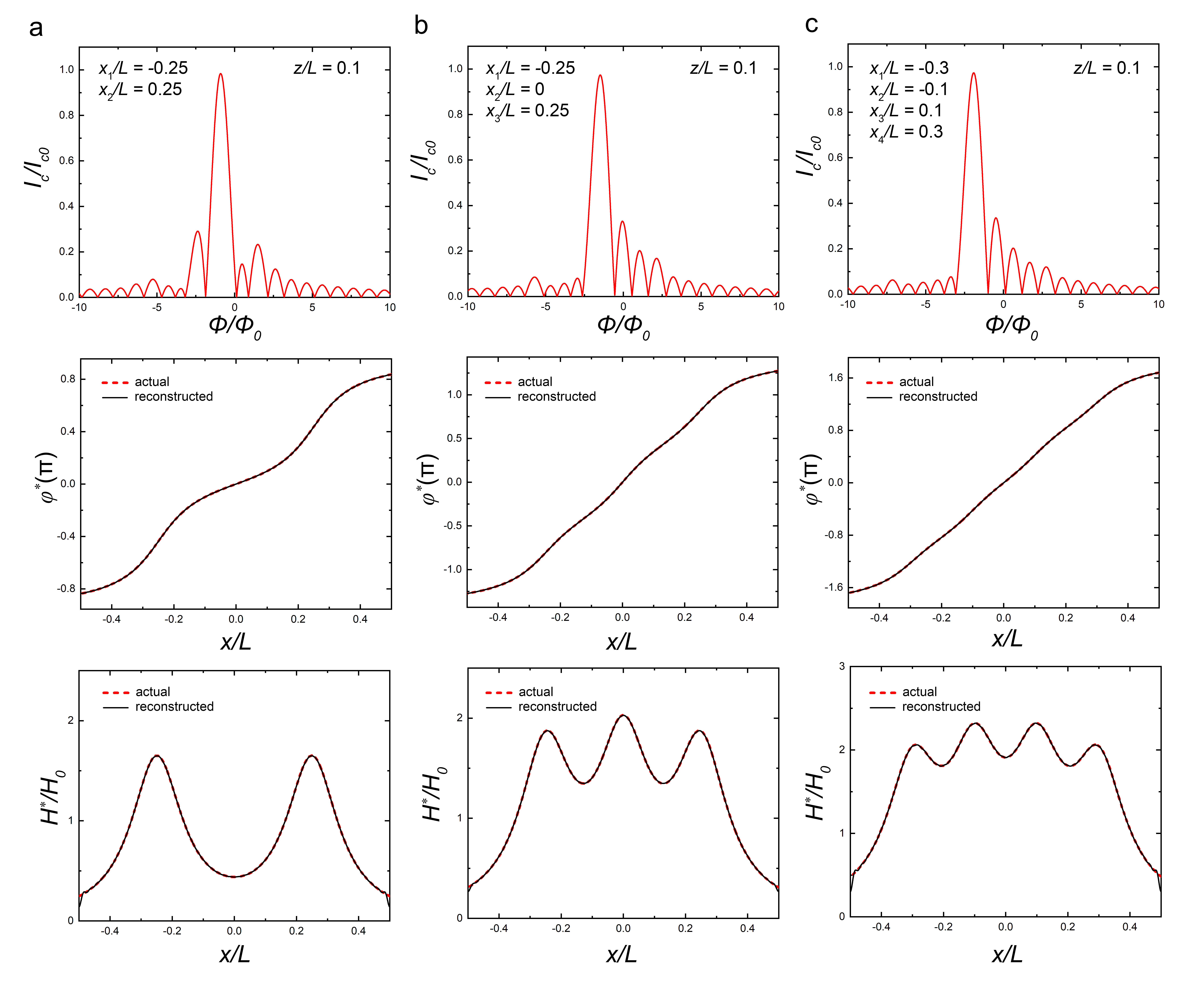}
        \caption{Examples of reconstructions of stray field distributions from (a) two, (b) three and (c) four Abrikosov antivorices, placed symmetrically along the junction at a distance, $z_v=0.1~L$. Top panels show $I_c(\Phi)$ modulation patterns. Middle and bottom panel represent local phase and field distributions. Red dashed lines represent actual distributions obtained from Eq. (\ref{AQ1}). Black lines represent results of reconstruction from $I_c(\Phi)$ patterns in the top panels.}
        \label{figS2}
\end{figure*}

Our approach bares certain resemblance with the seminal work by Dynes and Fulton \cite{Dynes_1971}, in which the inverse Fourier transform of $I_c(H)$ was employed for determination of $J_c(x)$ distribution. 
However, there are several key differences. First of all, we solve a different problem. 
Dynes and Fulton considered inhomogeneous JJs in a uniform field, leading to the integrand, $J_c(x)\exp(h x+\varphi_0)$, in Eq.~\ref{i I_c} with an unknown prefactor $J_c(x)$. 
To the contrary, we consider JJs in inhomogeneous field, leading to the integrand, $J_c\exp[h x+\varphi_0+\varphi^*(x)]$, with unknown $\varphi^*(x)$ {\em under the exponent}. Mathematically this is a different problem, which is more complicated due to its essential nonlinearity. It leads to a system of two Fourier integrals [our Eqs. (\ref{eqn6},\ref{eqn7})], as opposed to one (Eq. 3 in \cite{Dynes_1971}). 
Second, the aims of the two works are different. The objective of \cite{Dynes_1971} is to characterize the internal junction property, $J_c(x)$, while we are aiming to develop super-resolution magnetic imaging of external objects. 
For us the intrinsic $J_c(x)$ inhomogeneity is an unwanted artifact. Luckily, $J_c(x)$ cancels out upon division of Eqs (\ref{eqn6}) and (\ref{eqn7}) and does not hinder the reconstruction. 

\begin{figure*}[t]
        \includegraphics[width=0.99\textwidth]{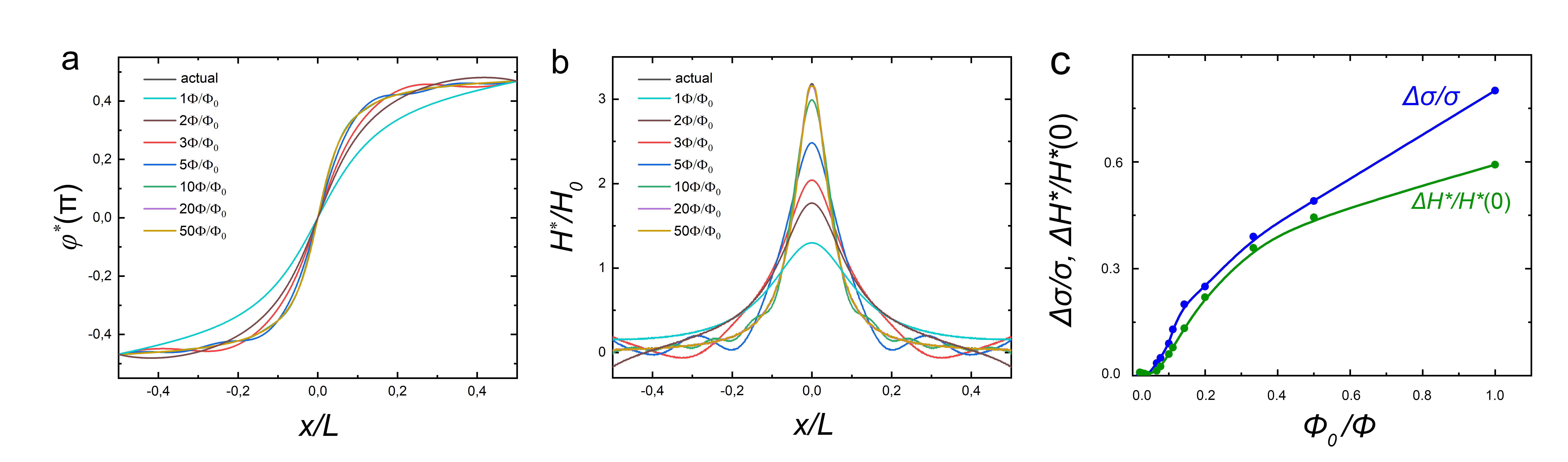}
        \caption{Development of image reconstruction for the case of Fig.~\ref{fig2}(c) upon successive truncation of $I_c(\Phi)$. (a) $\varphi^*(x)$ and (b) $H^*(x)$ obtained by integration of Eqs. (\ref{eqn6}) and (\ref{eqn7}) in a limited flux range $[-\Phi_{max},\Phi_{max}]$ with the maximum flux range $\Phi_{max}/\Phi_0$= 1, 2, 3, 5, 10, 20 and 50. Black lines in (a) and (b) represent the actual dependencies. (c) The relative accuracy of reconstruction versus the inverse flux range $\Phi_0/\Phi_{max}$. Blue and olive symbols show the relative errors of the width at half-maximum, $\Delta \sigma/\sigma$, and the height of the maximum $\Delta H^*/H^*(0)$, respectively, normalized by the actual values $\sigma$ and $H^*(0)$. It can be seen that the accuracy of reconstruction rapidly improves for $\Phi_{max}/\Phi_0>5$ and that the error practically vanishes at $\Phi_{max}/\Phi_0>10$.}
        \label{fig3n}
\end{figure*}

In what follows we restrict ourselves to the simplest case with symmetric local fields. This is done for two reasons. First, because it leads to a very simple mathematics, which is pedagogical for the proof-of-concept. We do confirm the existence of a more general solution for an arbitrary asymmetric case, but we leave it for a later occasion in a view of its complexity. Note, however, that the described simple procedure can be applied to any symmetric case because the integrand in $A(H)$ from Eq.~(\ref{f0}) remains odd with respect to $x$, leading to $A(H)=0$ and $\varphi_0=\pi/2$. 
The second reason is that the symmetric case is relevant for SPM, which commonly deals with small separated objects. To achieve it the sensor should be centered on top of the object. This can be easily done by maximization of the detected flux (see e.g. Fig. 3 (a) in Ref. \cite{Golod_2019a}). 

\subsection{Numerical verification}

To verify the method, first we consider the well calibrated case of AV. Vortex stray fields induce the Josephson phase shift \cite{Golod_2010,Golod_2019b}, \begin{equation}\label{AQ1}
\varphi^*(x) = -V \arctan \left(\frac{x-x_v}{|z_v|}\right).
\end{equation}
Here $V$ is the vorticity, and $x_v, z_v$ are AV coordinates. When the vortex approaches the JJ along the middle line, $x=0$, the total phase shift $\Delta\varphi^*=\varphi^*(L/2)-\varphi^*(-L/2)$, and the induced flux, $\Delta\Phi^*=(\Delta\varphi^*/2\pi)\Phi_0$, increase. This is shown by red dashed lines in the middle panels of Fig.~\ref{fig2}, which are calculated from Eq.~\ref{AQ1} with $V=-1$ (antivortex) and for (a) $z_v=\infty$, (b) $z_v=0.5~L$, and (c) $z_v=0.05~L$. Red lines in the top panels represent the direct problem solution: calculated $I_c(H)$ modulation, Eq.~\ref{I_c}, for given $\varphi^*(x)$ (red dashed lines in the middle panels). It is seen that the increase of $\Delta\varphi^*$ upon approaching the vortex to the junction leads to a progressive shift and distortion of $I_c(H)$ patterns \cite{Golod_2010,Golod_2019b,krasnov2020josephson,Golod_2021}. 
Solid black lines in middle and bottom panels of Fig. \ref{fig2} (a-c) represent the inverse problem solutions, $\varphi^*(x)$ and $H^*(x)$, reconstructed from $I_c(H)$ patterns from the top panels. They coincide with the actual profiles, shown by red dashed lines, confirming the successful image reconstruction. 

As mentioned above, $\varphi_0 =\pi/2$, remains well defined for any symmetric $H^*(x)$ allowing a straight forward integration of Eqs.~\ref{eqn6} and \ref{eqn7}. This facilitates unambiguous reconstruction of more complex multi-peak states. In Figure \ref{figS2} we demonstrate this for multivortex states with (a) two, (b) three and (c) four vortices placed symmetrically along the junction length. 

\subsection{Accuracy of reconstruction}

Local field reconstructions, presented in Figs.~\ref{fig2} and ~\ref{figS2}, are essentially perfect. They are obtained by integration of Eqs. (6,7) in the flux interval $[-50\Phi_0,50\Phi_0]$, i.e., from the $I_c(H)$ patterns with $\sim 100$ lobes. Of course, in practical cases the $I_c(H)$ pattern is usually measured in a narrower limit. In Fig. \ref{fig3n} we demonstrate how the reconstructed pattern is deteriorated upon successive truncation of the integration range $[-\Phi_{max},\Phi_{max}]$ for the case of Fig.~\ref{fig2} (c). Black lines in (a) and (b) represent actual $\varphi^*(x)$ and $H^*(x)$. The accuracy of $H^*(x)$ reconstruction in Fig. \ref{fig3n} (b) can be quantified by analysing the height of the maximum, $H^*(0)$, and the width at half-maximum, $\sigma$. Fig. \ref{fig3n} (c) shows corresponding relative errors as a function of the inverse flux range $\Phi_0/\Phi_{max}$. Here the olive symbols represent the relative error of the height, $\Delta H^*/H^*(0) = [H^*(0,\Phi_{max})-H^*(0,\infty)]/H^*(0,\infty)$, and the blue symbols - the relative error of the width, $\Delta \sigma/\sigma = [\sigma(\Phi_{max})-\sigma(\infty)]/\sigma(\infty)$, with $\sigma(\infty)$ and $H^*(0,\infty)$ being the actual parameters, corresponding to the black curve in Fig. \ref{fig3n} (b). It can be seen that the accuracy of reconstruction rapidly improves for $\Phi_{max}/\Phi_0>5$ and that the error practically vanishes at $\Phi_{max}/\Phi_0>10$.

From Figs. \ref{fig3n} (a) and (b) it can be seen that truncation of inverse Fourier integrals Eqs. (\ref{eqn6},\ref{eqn7}) lead to spurious oscillations, which have the wavelength $\lambda=(\Phi_0/\Phi_{max})L$. In the Appendix B we discuss two ways of image improvements using analytic continuation of truncated $I_c(H)$ patterns and Fourier filtering of spurious oscillations. 

\begin{figure*}[t]
    \begin{center}
        \includegraphics[width=0.95\textwidth]{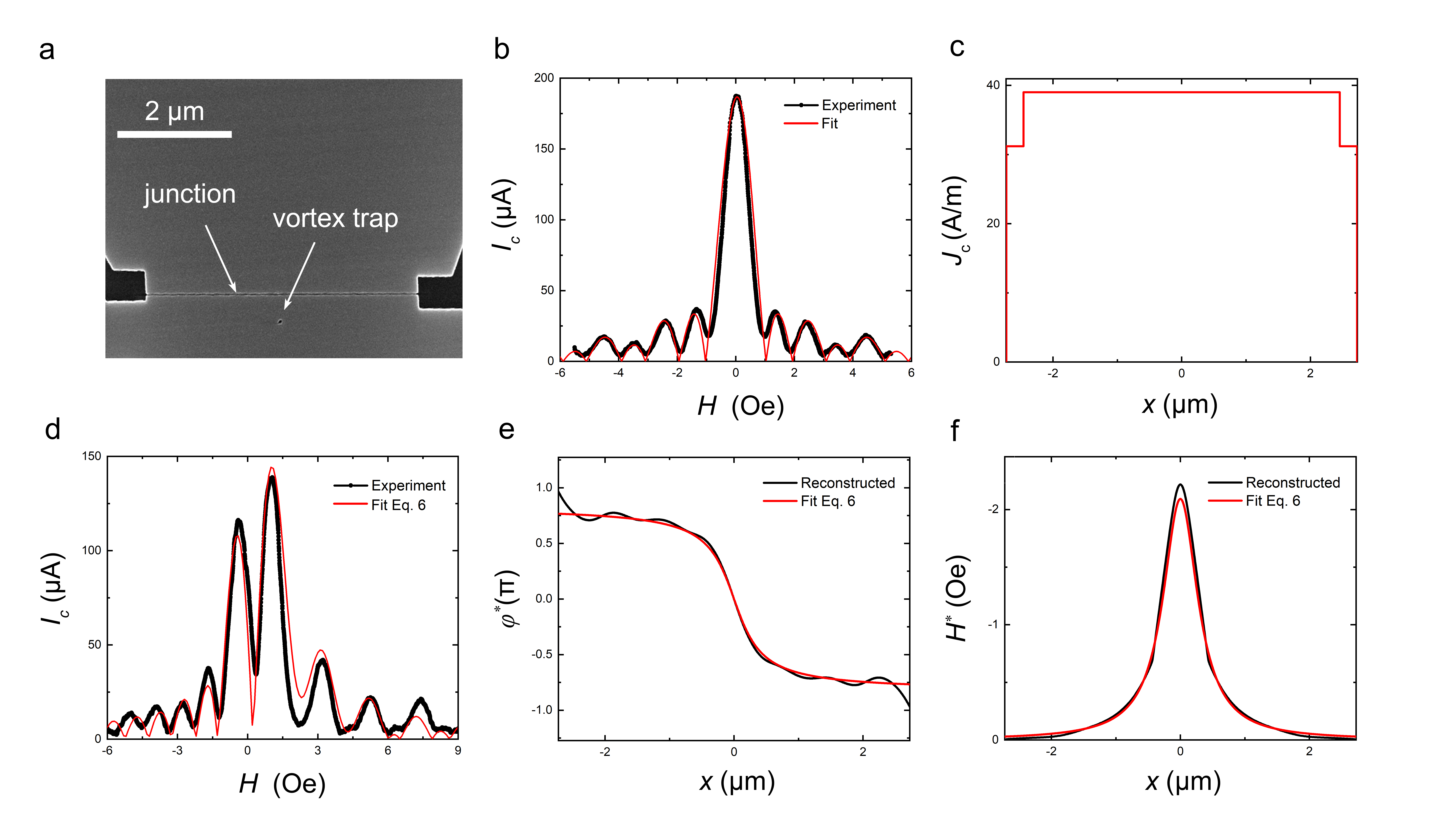}
        \caption{(a) SEM image of a planar Nb-CuNi-Nb junction with a vortex trap. (b) $I_c(H)$
pattern of the junction in the vortex-free case. Black symbols represent experimental data,
the red line is a fit, using the step-like $J_c(x)$ distribution, shown in (c). (d) Measured $I_c(H)$
pattern with a trapped vortex. Black symbols represent experimental data, the red line
is the fit, using Eq. (\ref{AQ1}). (e) and (f) show spatial distributions of (e) the phase shift and
(f) the vortex stray fields. Back lines represent the holographic reconstruction from the
experimental $I_c(H)$ pattern from (d). Red lines in (e) and (f) show expected distributions
obtained from the fit by Eq. (\ref{AQ1}) in (d).}
        \label{fig3}
    \end{center}
\end{figure*}

Our method resembles the Fourier-transform holography \cite{Bruckstein_1997,Latychevskaia_2015,Vetal_2018}, with diffraction-like $I_c(H)$ patterns serving as holograms. In holography the image quality increases with increasing the size of the hologram, i.e., with increasing the number of stored interference fringes. In our case the number of fringes corresponds to the number of lobes, i.e., to the flux range $\Phi_{max}/\Phi_0$. 
However, the specifics of our case is that the hologram is created by interference of the object with electronic wave functions of the condensate. In this respect it has a connection with electronic quantum holography \cite{Moon_2009}, which, however, occurs at a macroscopic scale in superconductors. 

\subsection{Experimental verification}.

For experimental verification we use planar Nb-CuNi-Nb JJs. Figure \ref{fig3} (a) show scanning electron microscope (SEM) image of the sample. Several devices were studies, each containing one or two JJs with the lengths $L\simeq 5.4~\mu$m and a vortex trap in the middle of the electrode, $x_v=0$, at different distances, $z_v$, from the JJs. For this sample, $z_v=0.36~\mu$m.
Variable thickness-type planar junctions are made by cutting CuNi(50 nm)/Nb(70 nm) bilayers by a focused ion beam (FIB). The bilayer is deposited by magnetron sputtering. Films are first patterned into $L \simeq 6~ \mu$m wide bridges by photolithography and reactive ion etching and subsequently cut by FIB to create JJs. Finally, a vortex trap (a hole with diameter $\sim 50$ nm) is made by FIB. Measurements are done in a closed-cycle $^4$He cryostat. The field is applied perpendicular to the junction plane. Details about device fabrication, characterization and the experimental setup can be found in Refs. \cite{Golod_2010,Boris_2013,Golod_2019a,Golod_2019b}. 

Black symbols in Fig. \ref{fig3} (b) show measured $I_c(H)$ patterns at $T\simeq 6.6$ K, in the absence of a vortex. It has a regular Fraunhofer-like shape with some minor beatings indicating step-like inhomogeneity of the critical current density \cite{Krasnov_1997,Dynes_1971}. The red line in (b) represents a fit to the $I_c(H)$ pattern with a step-like
$J_c(x)$, shown in Fig. \ref{fig3} (c).

Black symbols in Fig. \ref{fig3} (d) show the measured $I_c(H)$ after trapping a vortex. The vortex is introduced by a
current pulse, as described in Refs. \cite{Golod_2010,Golod_2019b}. The red line represents a
fit (direct problem solution) using $\varphi^*(x)$ from Eq.~(\ref{AQ1}) with the actual $L$, $x_v$ and $z_v$ and the prefactor
$V$ as the only fitting parameter. Red lines in Figs.~\ref{fig3} (e) and (f) show the corresponding expected $\varphi^*(x)$ and
$H^*(x)$ obtained from such the fit. Black lines in Figs. ~\ref{fig3} (e) and (f) represent reconstructed profiles (inverse solutions) obtained from the experimental $I_c(H)$ pattern from (d). The quantitative agreement with anticipated profiles (red lines) is apparent. Small
oscillatory deviations are due to the limited field range of the experimental $I_c(H)$. From
Fig.~\ref{fig3} (f) it is seen that the width at half-maximum of the reconstructed vortex stray field
is $\sim 500$ nm $<0.1 ~L$, confirming the super-resolution ability of the method.


\section{Discussion}

Finally, we discuss advantages of the planar geometry. Although the holographic method is applicable to any type of JJs, good resolution requires 5-10 lobes of $I_c(H)$ and the field range $\pm 5-10~ H_0$. This field should be small enough to be noninvasive for both the object and the sensor. Therefore, JJs with a high field sensitivity (small $H_0$) are preferred. In this respect, planar JJs with inherently small $H_0$ \cite{Boris_2013,Golod_2019a} have a major advantage compared to conventional overlap JJs. For our JJs $H = 6-8 $ Oe, see Figs. \ref{fig3} (a-c), is sufficient for achieving nano-scale spatial accuracy. Furthermore, as demonstrated earlier, see Fig. 4 (e,f) in Ref. \cite{Golod_2019a}, the planar geometry allows simple implementation of a control line for producing homogeneous magnetic field locally in the JJ. This facilitates acquisition of many $I_c(H)$ lobes without disturbance of the object. The ultimate field resolution of such sensor is determined by the flux noise. For our JJs it is $\sim 10^{-7}\Phi_0/\sqrt{Hz}$ at $T=4.2$ K \cite{Golod_2019a}. Taking into account the flux quantization field $H_0\simeq 1$ Oe, it translates to the ultimate field sensitivity of $10^{-11}$Oe$/\sqrt{Hz}$.
It is remarkable that, contrary to conventional imaging techniques, 
which suffer from the trade-off problem between sensitivity and resolution, in the discussed holographic method the high field sensitivity is {\em accompanied} by the high spatial resolution.


\section{Conclusions}

To conclude we derived theoretically and verified experimentally a method of magnetic image reconstruction by a single Josephson junction. It resembles holography, with the diffraction-like $I_c(H)$ pattern serving as a hologram. 
The method allows super-resolution image reconstruction with nano-scale spatial resolution not limited by the junction size. Thus, it can obviate the trade-off problem between the field sensitivity and the spatial resolution, typical for many imaging techniques, which directly probe the total flux or field in a sensor. We demonstrated that utilization of planar Josephson junctions for such holographic imaging facilitates both high field sensitivity and high spatial resolution, which is beneficial for scanning probe microscopy. 


\begin{acknowledgements}
We are gratefully to V.V. Dremov, S.Yu. Grebenchuk and V.S. Stolyarov for stimulating discussions. 
\end{acknowledgements}

\section*{Appendix A. Alternation of the sign of $I_c(\Phi)$ modulation}

\begin{figure}
    \begin{center}
        \includegraphics[width=.9\columnwidth]{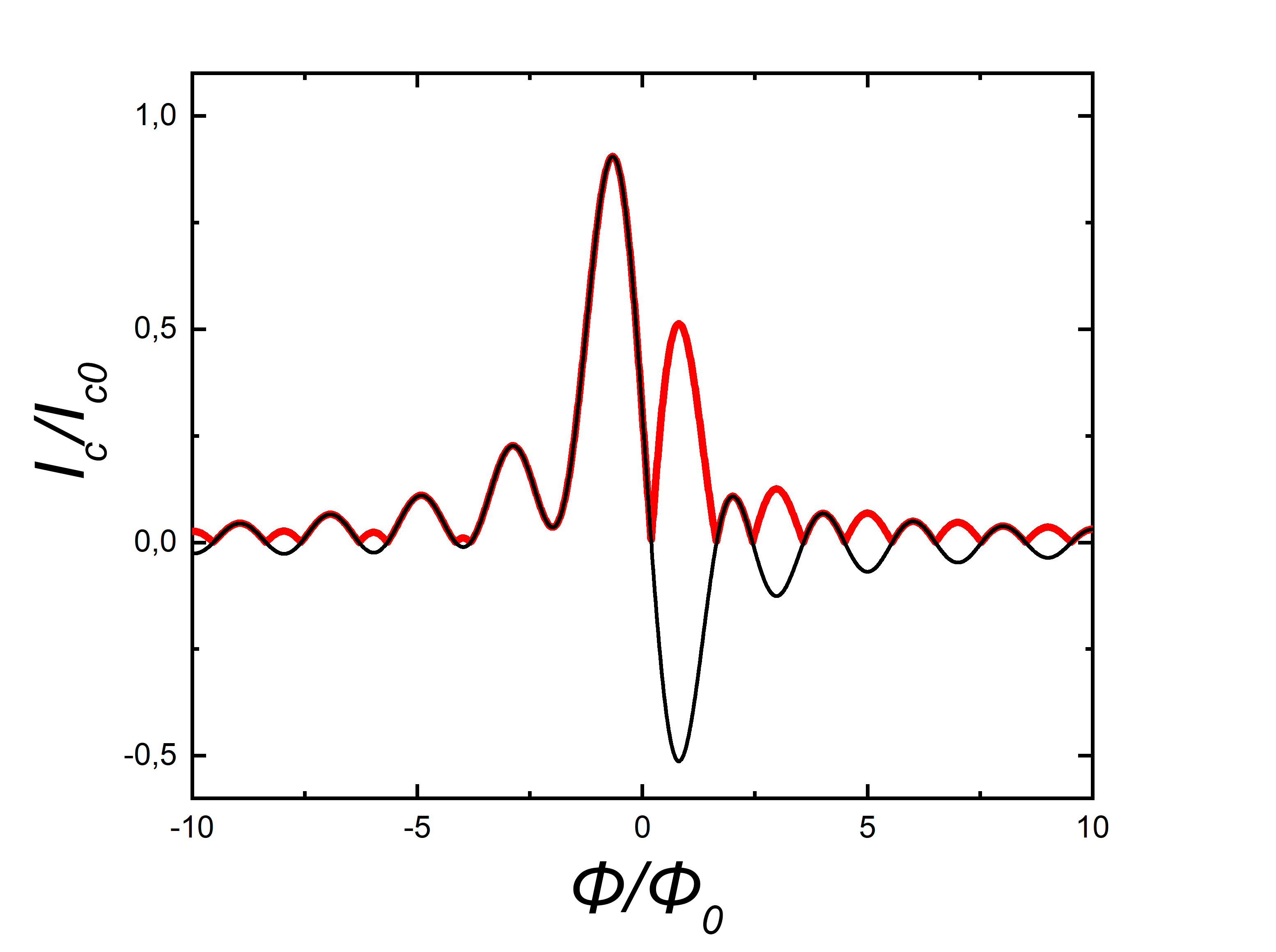}
        \caption{Demonstration of the sign-alternation procedure for the image reconstruction. The red line represents the $I_c(\Phi/\Phi_0)$ curve from Fig.~\ref{fig2}(c). The black line represents the same curve in which the sign of $I_c$ is sequentially flipped every time the $I_c(f)$ crosses zero. It is this black line that has to be used in the inverse Fourier integral, Eqs. (\ref{eqn6},\ref{eqn7}), with $\varphi_0=\pi/2$.}
        \label{figS1}
    \end{center}
\end{figure}

As described in Eq. (\ref{sign}), for  reconstruction with a fixed $\varphi_0=\pi/2$, the sign of the $I_c(\Phi)$ pattern has to be flipped every time the $I_c$ crosses zero. In Figure \ref{figS1} we show $I_c(\Phi/\Phi_0)$ modulation for the case from Fig. \ref{fig2} (c). Here the red line represents the ``experimental" $|I_{c}(\Phi/\Phi_0)|$ pattern and the black line - the curve with sequentially flipped sign, which has to be used for reconstruction via Eqs. (\ref{eqn6},\ref{eqn7}). It is seen that flipping of the sign should be made only when the $I_c(\Phi/\Phi_0)$ vanishes. For example, on the negative side the first maximum does not fall to zero and, therefore, the sign of the next lobe is preserved. Such flipping procedure has to be done in all cases prior to inverse Fourier transform.   

Experimental $I_c(H)$ characteristics can deviate from the ideal Fraunhofer modulation. In particular, from Fig. ~\ref{fig3}(b,d) it can be seen that the $I_c$ does not completely vanish at minima. Although, the non-vanishing $I_c$ can be caused by inhomogeneity~\cite{Dynes_1971}, we believe that the primary reason in our case is the finite length of the JJ with respect to $\lambda_J$. This follows from the temperature dependence of $I_c(H)$, studied in Ref.~\cite{Golod_2019a} for a similar JJ: The lower is $T$, the shorter is $\lambda_J$, the larger is the relative junction length, $L/\lambda_J$, and the more non-vanishing are $I_c(H)$ minima. 
To avoid this one should use short JJs with smaller $I_c$ and larger $\lambda_J$. However, in the pristine case, $H^*=0$, the $I_c$ flipping procedure remains unambiguous even for non-vanishing $I_c$: the $I_c$ should be flipped between odd-even lobes. Fortunately, the image reconstruction is not sensitive to a modestly non-vanishing $I_c$, as demonstrated by the successful reconstruction in Fig.~\ref{fig3}(e,f).   

Note that, Dynes and Fulton \cite{Dynes_1971} instead flipped $\varphi_0$, preserving the sign of $I_c$ (the saw-tooth like dependence $\Theta(H)$ in their case corresponds to the step-like $\pm \pi/2$ variation in our case because their $\Theta$ is determined at the edge of the JJ, $x=-L/2$). The two approaches, flipping $I_c$ or $\varphi_0$, are identical. We have chosen to flip $I_c$ because it allows explicit operation with integral sine a cosine functions, as described in appendix B.

\section*{Appendix B. Image improvement methods}

The integral in Eqs. (5-7) is taken in the infinite field range. However, in experimental situation the field range is always finite. This inevitably leads to distortions in the reconstructed image. As can be seen from Fig.~\ref{fig3n} (b), the agreement with the actual $\varphi^*(x)$ variation (black line) becomes satisfactory only when there are at least five lobes in $I_c(\Phi)$ for each field direction. For more than ten lobes the agreement becomes very good, but this is not always feasible in experiment. Therefore, methods for improving image reconstruction from truncated $I_c(H)$ pattern are required. We tested two such methods, as described below. 

\subsection*{Analytic continuation}
\begin{figure*}
   \begin{center}
        \includegraphics[width=0.8\textwidth]{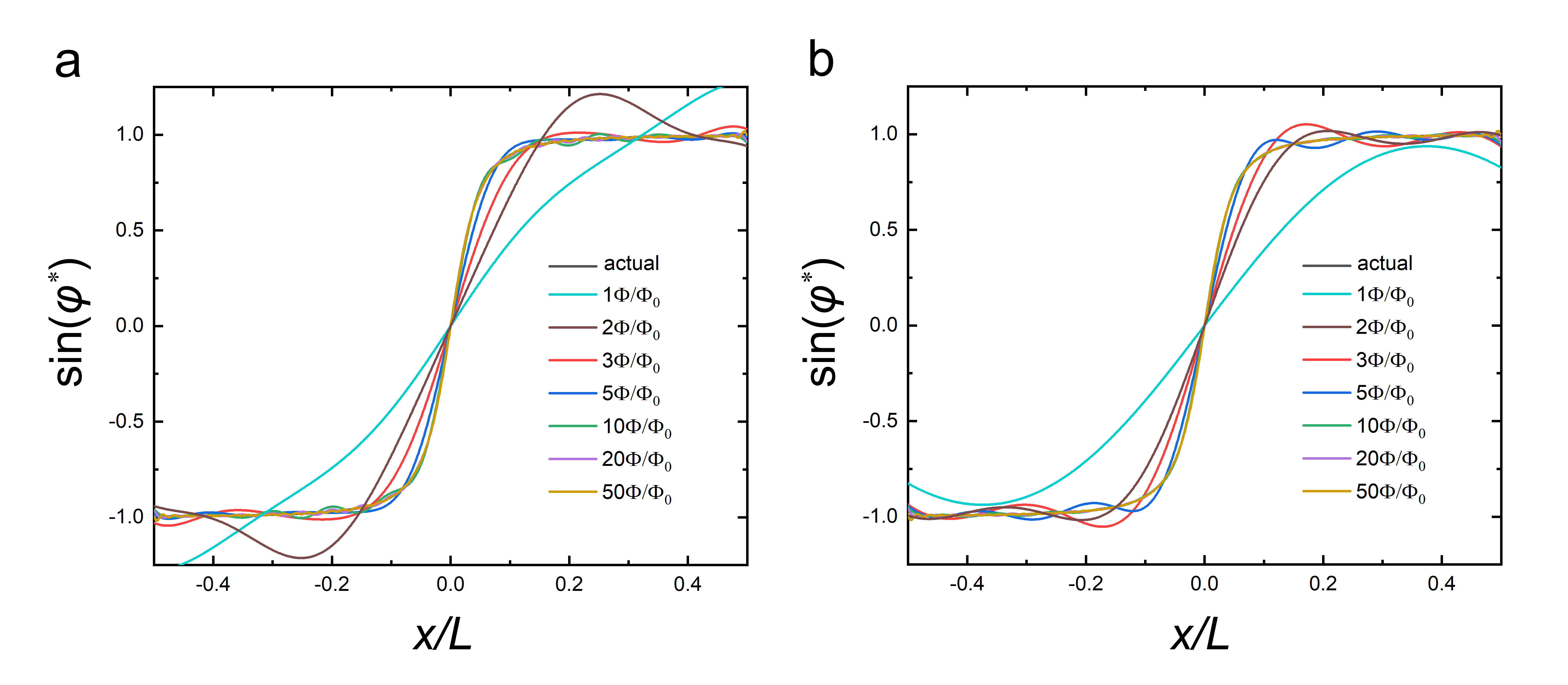}
        \caption{Demonstration of improvements via analytic continuation of the same truncated $I_c(\Phi)$ patterns as in Fig.~\ref{fig3n}. Curves in (a) are obtained using analyric continuation, Eq. (\ref{EQ4}) in flux range $\Phi_{max}=-\Phi_- = \Phi_+$ with cut-off at $\Phi_{max}/\Phi_0$= 1, 2, 3, 5, 10, 20 and 50. Curves in (c) are obtained by symmetric truncation with respect to the central maximum of $I_c(\Phi)$, i.e., in the flux range $\Phi_{max}=-\Phi_- + \Phi^*=\Phi_+ + \Phi^*$, using the continuation Eq. (\ref{SimAdded}). It can be seen that both types of analytic continuation improve image reconstruction with a certain advantage for symmetric truncation with respect to the central peak (b). 
        }
        \label{Opt}
    \end{center}
\end{figure*}
The continuation method is aiming to add an analytic continuation to the truncated $I_c(\Phi)$. At large fields $H\gg H^*$, the effect of local field becomes insignificant and the $I_c(\Phi)$ modulation resumes the Fraunhofer shape. However, since the local field introduces a certain flux, $\Phi^*$, in the JJ, the $I_c(\Phi)$ modulation is shifted by $-\Phi^*$. Therefore, the truncated pattern must be complemented by the shifted Fraunhofer function $I_c = \bigg|\frac{\sin[\pi (\Phi+\Phi^*)/\Phi_0]}{[\pi (\Phi+\Phi^*)/\Phi_0]} \bigg|$. Here $\Phi=\beta L \Phi_0 H$ is the flux induced by applied field $H$. If the truncated $I_c(H)$ pattern is measured in the interval $[H_-,H_+]$ then
\begin{figure*}[!ht]
   \begin{center}
        \includegraphics[width=1.99\columnwidth]{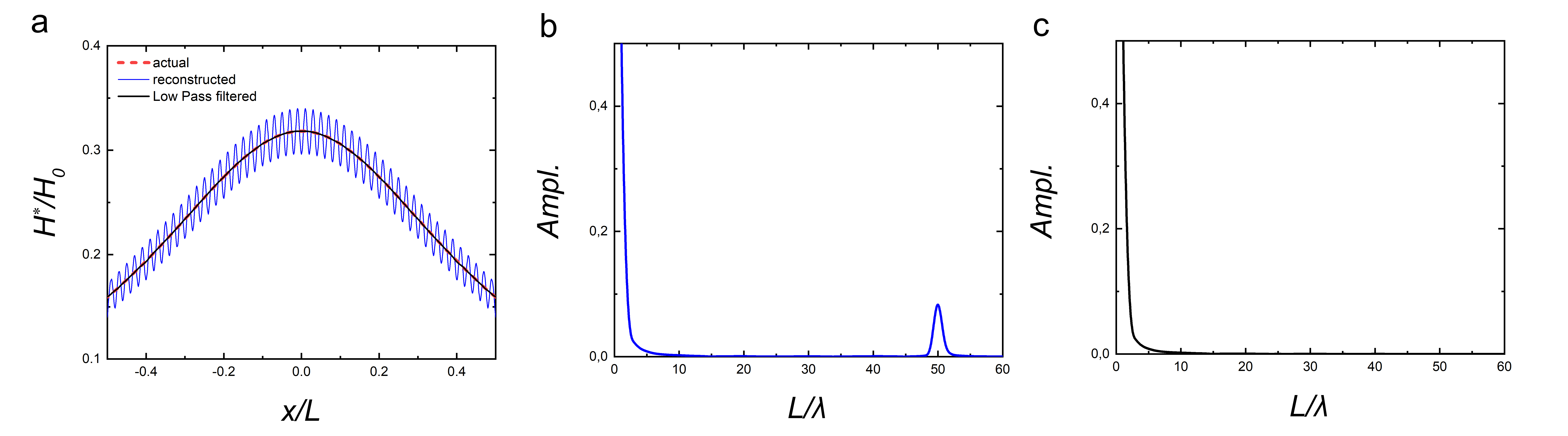}
        \caption{(a) Reconstructed field for the case of Fig. ~\ref{fig2}(b)  middle. The red dashed line shows the actual $H^*(x)$. The blue line shows directly reconstructed image from $I_c(\Phi)$ truncated at $\Phi_{max}/\Phi_0=50$. Short wavelength oscillations, caused by the truncation, are clearly seen. The black line shown the same curve after Fourier filtering with a with low pass filtering of parasitic oscillations. (b) FFT spectrum of the blue curve from (a). A small maximum can be seen at $L/\lambda=\Phi_{max}/\Phi_0=50$ (c) The low pass filtered spectrum, used for reconstruction of the black curve in (a). }
        \label{Fur}
    \end{center}
\end{figure*}

\begin{widetext}
\begin{equation}
\label{Added}
   \begin{split}
J_c(x) \sin[\varphi^*(x)] =\beta \int\limits_{H_-}^{H_+} \cos \bigg(\alpha x H + \varphi_0(H)\bigg)I_c(H)dH + \frac{1}{2\pi}f(x). 
   \end{split}
\end{equation}
\end{widetext}
For $\Phi^* < 0$ and $\Phi_+ > \Phi^* $, 
the complementary function is equal to 
\begin{widetext}
    \begin{equation}
    \label{EQ4}
    \begin{split}
            f(x) = \sin\left(2\pi x\frac{\Phi^*}{L\Phi^0} \right) \bigg(2\pi -\mathrm{Si}\bigg(\pi(1-2x)\frac{|\Phi_- + \Phi^*|}{L\Phi_0}\bigg) - \mathrm{Si}\bigg(\pi(1+2x)\frac{|\Phi_- + \Phi^*|}{L\Phi_0}\bigg)
            - \mathrm{Si}\bigg(\pi(1-2x)\frac{\Phi_+ + \Phi^*}{L\Phi_0}\bigg) - \\ - \mathrm{Si}\bigg(\pi(1+2x)\frac{\Phi_+ + \Phi^*}{L\Phi_0}\bigg)\bigg) +
            \cos\left(2\pi x\frac{\Phi^*}{L\Phi^0} \right)\bigg(\mathrm{Ci}\bigg(\pi(1+2x)\frac{|\Phi_- + \Phi^*|}{L\Phi_0}\bigg) - \mathrm{Ci}\bigg(\pi(1-2x)\frac{|\Phi_- + \Phi^*|}{L\Phi_0}\bigg) - \\ -\mathrm{Ci}\bigg(\pi(1+2x)\frac{\Phi_+ + \Phi^*}{L\Phi_0}\bigg)  + \mathrm{Ci}\bigg(\pi(1-2x)\frac{\Phi_+ + \Phi^*}{L\Phi_0}\bigg)\bigg).
        \end{split}
    \end{equation}
\end{widetext}

Here $\mathrm{Si}$ and $\mathrm{Ci}$ are integral sine and cosine functions, respectively.
Eq. (\ref{EQ4}) may contain singularity points because 
$\lim_{x \to 0} \mathrm{Ci}(x)  \rightarrow  \infty$. To avoid problems associated with the singularity it is advisable to introuce a symmetric truncation with respect to the central maximum at $\Phi=-\Phi^*$: $|\Phi_- +\Phi^*|=|\Phi_++\Phi^*|$. Furthermore, this makes sense because the major part of information about $H^*(x)$ is concentrated around the central maximum. In this case the complimentary term in Eq.~(\ref{EQ4}) is simplified to  
\begin{widetext}
    \begin{equation}
    \label{SimAdded}
    \begin{split}
           f_{sym}(x) = \sin\left(2\pi x\frac{\Phi^*}{L\Phi_0} \right)  \bigg(2\pi - 2\mathrm{Si}\bigg(\pi(1-2x)\frac{\Phi_+ + \Phi^*}{L\Phi_0}\bigg) - 2\mathrm{Si}\bigg(\pi(1+2x)\frac{\Phi_+ + \Phi^*}{L\Phi_0}\bigg)\bigg).
        \end{split}
   \end{equation}
   \end{widetext}

 To investigate how the proposed continuation affect the quality of reconstruction, in Figs. ~\ref{Opt} (a) and (b) we show the reconstructed $\sin(\varphi^*)$ profiles for the same conditions as in Fig.~\ref{fig3n}, using analytic continuations (a) Eq. (\ref{Added}) and (b) Eq. (\ref{EQ4}). Note that the reconstruction significantly improved when $\varphi^*$ is calculated from $\tan\varphi^*$. For this the same method of continuation should be applied for $\cos \varphi^*$ as well. It can be seen that both types of continuation improve the reconstruction. However, a truncation symmetric with respect to the central maximum,  Eq.~(\ref{SimAdded}), provides the best result.

\subsection*{Fourier filtering}

 From Fig.~\ref{fig3n} (a,b) it is seen that truncation of $I_c(\Phi)$ patterns leads to the appearance of parasitic oscillations in the reconstructed image. The corresponding two wavelengths can be deduced from the correction function $f(x)$ in Eq. (12). The long wavelength, $\lambda_1 = L \Phi_0/\Phi^*$, is represented by the $\sin(2\pi x\Phi^*/\Phi_0 )$ term. The short wavelength,
$\lambda_2=L \Phi_0/(\Phi_{max}+\Phi^*)$, is associated with the $\mathrm{Si}$ terms. The wave number of the latter is proportional to the total flux interval. Such behavior can be seen in simulations shown in Fig. ~\ref{fig3n}(a,b), which also indicates that the short wavelength oscillations at the edges of the JJ remain even for large flux ranges. However, such shortwave oscillations can be very effectively removed by low pass or band block filtering. 

Figure ~\ref{Fur}(a) shows the $H^*(x)$ dependency for the case of Fig.~\ref{fig2}(b) middle. Here the red dashed line shows the actual $H^*$ and the blue curve - directly reconstructed, using Eq. (\ref{eqn6},\ref{eqn7}) for $\Phi_{max}/\Phi_0=50$. Parasitic short wavelength oscillations are clearly seen. Fig. 6(b) shows the Fourier spectrum of the reconstructed image. A peak at $L/\lambda=\Phi_{max}/\Phi_0=50$ is clearly seen. Since it is well separated from the low wave number part, which represents the actual $H^*(x)$ variation, it can be simply removed by proper band pass filtering. The result of such filtering is shown in Fig.~\ref{Fur}(c). The black line in Fig.~\ref{Fur}(a) shows the $H^*(x)$ profile obtained from the Fourier filtered spectrum from (c). The agreement with the actual dependence (dashed red line) is perfect. The success of Fourier filtering method depends on the spectral separation of the informative peak at $L/\lambda \rightarrow 0$ and the artifact peak at $L/\lambda=\Phi_{max}/\Phi_0$. This again requires large enough flux range with $\Phi_{max}/\Phi_0>5$.\\





\end{document}